\documentclass[twoside,11pt]{article}
\usepackage{journal, theapa, rawfonts}
\usepackage{amsmath,amssymb,amsfonts}
\usepackage{algorithmic}
\usepackage{graphicx}
\usepackage{amsmath,amssymb,amsfonts}
\usepackage{caption}
\usepackage{subcaption}
\usepackage{stackengine}
\usepackage{mathtools}
\usepackage{float}
\usepackage[shortlabels]{enumitem}
\usepackage{url} 
 
\begin{document}

\title{Towards a Semantic Information Theory \\(Introducing Quantum Corollas)}

\author{\name Philip Tetlow \email philip.tetlow@uk.ibm.com \\
       \addr IBM Global Markets, Leeds, United Kingdom
       \AND
       \name Dinesh Garg \email garg.dinesh@in.ibm.com \\
       \addr IBM Research, Bangalore, India
       \AND
       \name Leigh Chase \email leigh\_chase@uk.ibm.com \\
       \addr IBM Enterprise \& Technology Security, Hursley, United Kingdom
       \AND
       \name Mark Mattingley-Scott\thanks{The work was done while the author was working for IBM Research, Germany.}  \email mark.mattingly-scott@quantum-brilliance.com\\
       \addr Quantum Brilliance, Stuttgart, Germany
       \AND
       \name Nicholas Bronn \email ntbronn@us.ibm.com \\
       \addr IBM Research, Yorktown Heights, United States
       \AND
       \name Kugendran Naidoo\thanks{The work was done while the author was working for IBM Research, South Africa.}  \email k@qtda.net \\
       \addr Independent, Johannesburg, South Africa
       \AND
       \name Emil Reinert \email Emil.Reinert@ibm.com \\
       \addr IBM Global Markets, Aarhus, Denmark
       }

% For research notes, remove the comment character in the line below.
% \researchnote

\maketitle

\begin{abstract}
The field of Information Theory is founded on Claude Shannon's seminal ideas relating to entropy. Nevertheless, his well-known avoidance of meaning \cite{shannon48} still persists to this day, so that Information Theory remains poorly connected to many fields with clear informational content and a dependence on semantics. Herein we propose an extension to Quantum Information Theory which, subject to constraints, applies quantum entanglement and information entropy as linguistic tools that model semantics through measures of both \textit{difference} and \textit{equivalence}. This extension integrates Denotational Semantics with Information Theory via a model based on distributional representation and partial data triples known as Corolla.
\end{abstract}

\section{Introduction}
\label{Introduction}

Established research suggests that standard approaches to quantum computing can model the basic concepts of formal semantics \cite{widdows20} \cite{smolenksy90} \cite{kanerva09} \cite{clark07} \cite{widdows08} \cite{mitchell10} \cite{baroni10} \cite{turney12} \cite{garg19} \cite{cohen12} \cite{cohen11}. Furthermore, fields like Information Retrieval (IR) have long used vectors and linear algebra to model, manipulate and interrogate complex bodies of information \cite{widdows20} \cite{rijsbergen04} \cite{widdows04}. This \textit{quantum-inspired} work \cite{khrennikov10} has introduced parallels with quantum mechanics, and therein Quantum Computing (QC), which remove many practical distinctions between the two fields \cite{rijsbergen04} \cite{widdows04}. For that reason, the idea of exploiting information en masse using quantum methods is highly appealing but remains marred by the practicalities of physical implementation. It is also complicated by at least one duality at the crossover between mathematics, science, computing and information theory. This concerns the notions of abstraction, primitive definition, and axioms and can be explained directly in terms of vectors themselves.

%-------------------------------------------
\begin{figure}
\centering
\includegraphics[width=0.3\columnwidth]{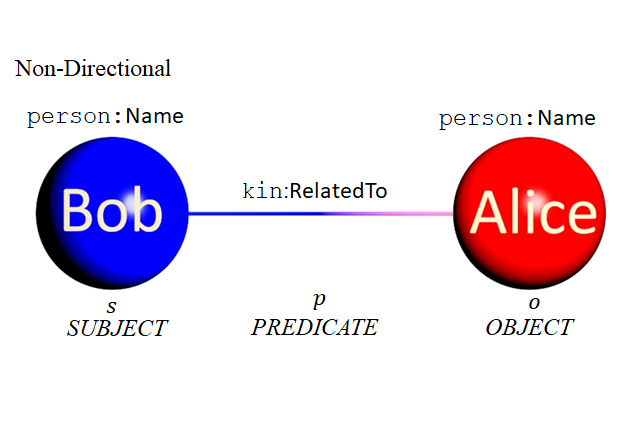}
\includegraphics[width=0.3\columnwidth]{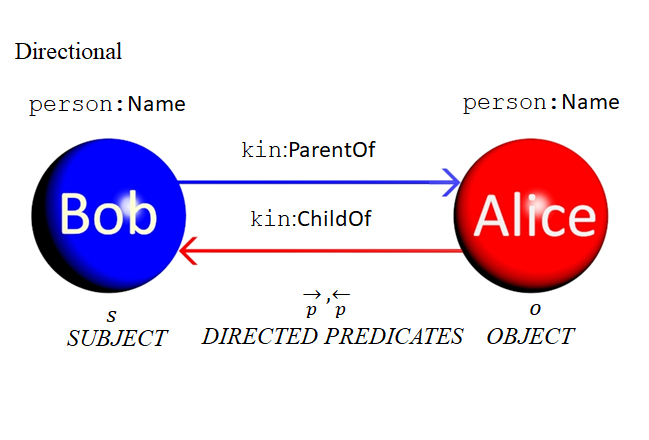}
\includegraphics[width=0.3\columnwidth]{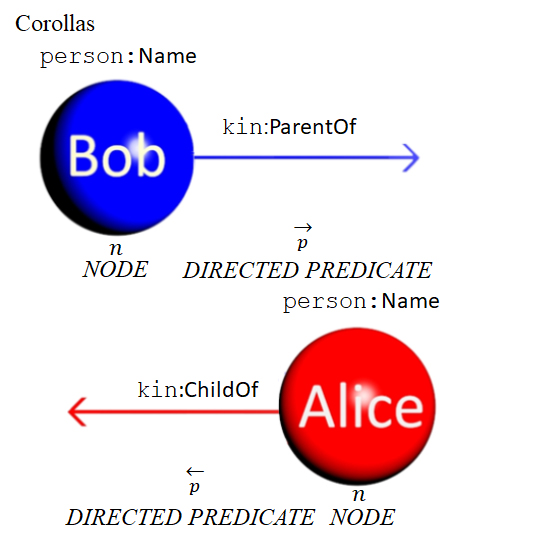}
\caption{Non-directional, Directional and Corolla Examples of a Semantic Triple}
\label{Figure 1(a), 1(b) and 1(c)}
\end{figure}
%-------------------------------------------

A vector, as a structured mathematical object having magnitude and direction within some abstract space, can represent some set of characteristics sufficient to describe a \textit{pure} \textit{state} system. In that regard, it presents as a primitive, requiring no further justification or decomposition other than through the dimensions (bases) and scalars that aid qualification and quantification. Yet vectors are constructed from a potentially infinite set of discrete points along their projection, so they are equally nonprimitive. Therefore, by definition, they are also (potentially) described by at least one mixed state \cite{widdows20} - a probabilistic weighted sum of outer vector products, which in quantum mechanics is referred to as a \textit{density} \textit{matrix}. Consequently, to gain a complete understanding of any system described using vectors, one needs to appreciate both primitive and non-primitive interpretations. To make this understanding concrete, the definition of any truly primitive parts must be open to derivation from first principles: that is, they should be provably axiomatic so that any further decomposition would render their justification \textit{usefully} \textit{meaningless}. In turn, that requires the precise definition of any bases and interval(s) over which individual primitives might be defined and the limits that close them.
\
Therefore, this paper proposes a novel way to achieve axiomatic representation and composition in vector-based information systems using the partial-triple construct known as Corolla. This formally defines primitives in terms of the denotational semantics they cover and those of the immediately relevant directed relationships they offer.

\section{Towards Quantum Semantics}
\label{quantum_semantics}

One can consider logical non-directional semantic triples as a formal specification of a conceptualisation in the form $(s, p, o)$; where $s$ and $o$ represent subject and object concepts, entities or strings, and $p$ a non-directional predicate relationship between them. An example is shown in Figure \ref{Figure 1(a), 1(b) and 1(c)}, explaining that {\em Bob and Alice are related as kin}, and by adding predicate direction, all such triples expand to entail 
$\left( s, \text{\stackanchor{$\rightarrow$}{$p$}}, \text{\stackanchor{$\leftarrow$}{$p$}}, o \right)$ meaning that two directed predicate (bilinear) relationships result from a single non-directional relationship. 

In the example shown, this first gives the directed triple (\texttt{person}:{\bf Bob}, \texttt{kin}:{\bf ParentOf}, \texttt{person}:{\bf Alice}), making explicit the entailment \texttt{kin}:{\bf ParentOf}, from the perspective of \texttt{person}:{\bf Bob} as the triple’s subject, and models the literal meaning {\em Bob is the father of Alice} -- where $s$ and $o$ are semantically grounded to represent instances of type \texttt{person}, with assigned values of {\bf Bob} and {\bf Alice}, as shown in Figure \ref{Figure 1(a), 1(b) and 1(c)}. Likewise, the directed predicate
\stackanchor{$\rightarrow$}{$p$} is semantically grounded to an instance of type \texttt{kin} with a value of {\bf parentOf}. Second, the triple’s converse meaning logically instantiates as (\texttt{person}:{\bf Alice}, \texttt{kin}:{\bf childOf}, \texttt{person}:{\bf Bob}), making explicit the entailment \texttt{kin}:{\bf ChildOf}, from the perspective of \texttt{person}:{\bf Alice}, literally meaning {\em Alice is the daughter of Bob} and involving the directed predicate \stackanchor{$\leftarrow$}{$p$} grounded to \texttt{kin:}C{\bf hildOf}. Similarly, both (\texttt{person}:{\bf Bob}, \texttt{kin}:{\bf HusbandOf}, \texttt{person}:{\bf Mary}) and (\texttt{person}:{\bf Mary}, \texttt{kin}:{\bf WifeOf}, \texttt{person}:{\bf Bob}) provide a comparable example explaining that Bob and Mary are a married couple. However, the key point in all of these cases is that the nodes involved contribute equally to the semantics of any nondirectional relationship between them. They therefore play a half-part in the meaning of any shared nondirectional predicate join.

By combining a triple node with a unary directed predicate as its owner, the notion of a Corolla is introduced, as in Figure \ref{Figure 1(a), 1(b) and 1(c)}(c), which corresponds to a half-part nondirectional triple. Consequently, corollas come in pairs so that, for instance, in the triple (\texttt{person}:{\bf Bob}, \texttt{kin}:{\bf ParentOf}, \texttt{person}:{\bf Alice}), (\texttt{person}:{\bf Bob}, \texttt{kin}:{\bf ParentOf}) provides one half of the triple’s corolla pair and (\texttt{person}:{\bf Alice}, \texttt{kin}:{\bf childOf}) its converse partner. Without loss of generality then, all corolla denote logically as $\left(n, p_{\text{directed}}\right)$, where $n$ represents the semantics assigned to any logical node within a semantic ontology or knowledge graph, and $p_{\text{directed}}$ the semantics assigned to any logical predicate that $n$ might legitimately use to form a directed predicate relationship with other nodes in the same graph. To use an analogy from the physical sciences, $n$ therefore corresponds to an instance of a known atom type and $p_{\text{directed}}$ the electrons donated or received by $n$ in the different chemical bonds that that atom might legitimately use when reacting with other atoms. 

\begin{figure}[H]
\centering
\includegraphics[width=0.8\columnwidth]{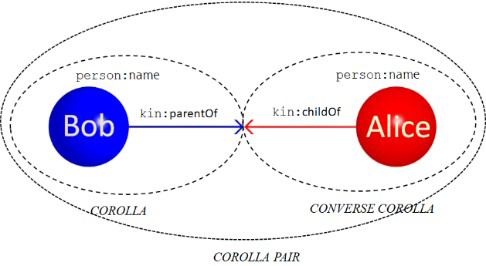}
\caption{An Example of a Corolla Pair Joining to make a Semantic Triple}
\label{Figure 2}
\end{figure}

This creates a model of a (directed) triple as a composite system comprising two distinct corollas $\left(c_i,c_j \right)$ as illustrated in the example shown in Figure \ref{Figure 2}; each corolla being drawn from a finite, countable set of all possible legitimate and unique alternatives:

An appropriate graph grammar \cite{marcolli15} captures this, as:
\begin{eqnarray*}
G &=& \text{Any given graph}\\
N(G) &=& \text{Finite set of all nodes within } G\\
C(G) &=& \text{Finite set of corollas with assigned values within } G\\
F(G) &=& \text{Finite set of all half-edges within any graph } G
\end{eqnarray*}
\begin{eqnarray}
\text{\bf Involution } && I : F(G) \mapsto F(G)\\
\text{\bf Edges} && (f, f^{'}) \text{ with } f \neq f^{'} \in F(G),  \text{ with } \\
&& I(f) = f^{'} \text{ i.e. converse semantics persist} \nonumber\\
&& \text{across half-edge pairs to make a full edge}\nonumber \\
\text{\bf Triples} && \left(c,c'\right), \text{ where } c \neq c' \in C(G) 	
\end{eqnarray}

%-------------------------------------------
%-------------------------------------------

\section{Prior Art and Physical Instantiation}
\label{prior_art}
Vector Symbolic Architectures (VSA) cover a family of related approaches that can be implemented as logical connectionist systems and share a commitment to algebraic operations on distributed representations over highly dimensional vector spaces \cite{gayler04} -  which makes them immediately relevant to quantum computing architectures. They are descended from early work \cite{smolenksy90} on tensor product variable binding networks that demonstrated the validity of  variable-to-value binding and the representation and manipulation of complex nested graph structures using connectionist methods.

VSA rely on algebraic operations with \textit{simple} connectionist implementations, with the consequence that full tensor product use becomes thoroughly impractical because the vector dimensionality increases exponentially with the depth of the structures to be represented. Physically, however, this challenge can be overcome when two or more quantum systems are manipulated into a partially entangled state, thereby restricting the range of relationship open to measurement.

VSA retain the advantages of tensor product binding while avoiding the problem of increasing vector dimensionality.  In tensor product binding, the representation of the association of two entities is created as the outer product of the vectors representing the two entities. Thus, if the entity vectors are of dimensionality $n$, the outer product will be of dimensionality $n^2$.  VSA overcome this problem of increasing dimensionality by applying a function to the $n^2$ elements of the outer product to yield a resultant vector of dimensionality $n$. Thus, all structures, whether atomic or complex, are represented by vectors of the same dimensionality. 

When considering ways to realise semantic triples in an actual quantum environment, it is important to understand the physical engineering challenges associated with VSA-like arrangements \cite{cohen11} \cite{gayler04}. These require the non-trivial scaling of quantum systems and their associations to provide the necessary volume of quantum resources, including entanglement \cite{schneeloch19}.

As quantum information practice matures, the need to separate the physical layer that provides the quantum resources (like superconducting circuits, trapped ions,nitrogen-vacancy centers and photons) from the logical layer that utilises these resources will become clear.  For instance, many imperfect quantum systems may form one logical unit of information, or thousands of atoms may coherently act as a unit of quantum memory.

Also, as with classical communication and computing, protocols and algorithms will be implemented in the logical layer with minimal concern for the underlying platform, irrespective of whether real-world systems are varied and imperfect \cite{schneeloch19}. Today this is seen by those developing algorithms for quantum computers, who must consider the layout (or topology) of the qubits within a given solution. This includes specialised implementations of multiplication-like linear operators to achieve primitive \textit{binding} and assumes that both vector product inversion and source primitive recovery can be practically achieved. That is to say, in accordance with established VSA practice, if \includegraphics[]{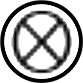} acts as a linear \textit{binding} operator and \includegraphics[]{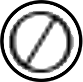} its linear \textit{unbinding} inversion, then if  $c$ = $a$ \includegraphics[]{img/oox.png} $b$, so $a$ \includegraphics[]{img/ool.png}  $c$ = $a$ \includegraphics[]{img/ool.png}  ($a$ \includegraphics[]{img/oox.png}  $b$) = $b$ \cite{cohen15} – accepting that in some contexts this recovery may be approximate. Thus, the \textit{bind} operator’s invertible nature facilitates the retrieval of information encoded during the \textit{binding} process. While this operator varies across VSA’s, it results in a product that is of the same dimensionality as the component vectors from which it was derived, unlike the tensor product, which has the dimensionality of its component vectors squared. When XOR is used, \textit{binding} commutes. Therefore $a$ \includegraphics[]{img/oox.png} $b$ = $b$ \includegraphics[]{img/oox.png} $a$ \cite{cohen15}, implying nondirectional binding. The use of standard tensor product operators does not commute and implies bilinear \textit{binding} in the form of bidirectional relations. It also cannot be inverted and separated into constituent primitives easily. But that does not mean that constituent primitives cannot be separated and accounted for individually by measurement when instantiated physically as quantum particles (hereafter referred to as quantum systems).

\section{Extending the Qudit Model for Generic Symbolic Representation}
According to standard practice, units of quantum information are normally considered by way of either qubit or qudit representation: where the assigned base system is used to delineate the symbols in use (normally “0” and “1” in the case of a qubit) and a qudit is defined (from computing) as the unit of quantum information described by a superposition across \textit{d} basis vectors and where \textit{d} is an integer greater than 2. Thus, qudits are described by a vector in a $d$ dimensional Hilbert space \(\mathfrak{H}_d\), spanned by a set of orthogonal basis vectors \{${|0\rangle,|1\rangle,|2\rangle,...|d-1\rangle}$\} and have the general form \cite{wang20}:\\
\begin{eqnarray}
    x = |\alpha\rangle=\alpha_0|0\rangle+\alpha_|1\rangle+\alpha_2|2\rangle ... + \alpha_{d-1}|d-1\rangle = \begin{pmatrix}
        \alpha_0 \\
        \alpha_1 \\
        \alpha_2 \\
        ... \\
        \alpha_{d-1}
        
      \end{pmatrix}  \ \in \mathbb{C}^d
\end{eqnarray}
and where: 
\begin{eqnarray}
    |\alpha_0|^2 + |\alpha_1|^2 + |\alpha_2|^2 ... + |\alpha_{d-1}|^2 = 1
\end{eqnarray}
\\

Assuming that the qudit model does not subsume that of the qubit, it is therefore not broad enough to map onto a complete symbolic view of information. Instead, a definition is needed that extends the qudit to include information representation where $0 < d > 3$; as is partially covered by standard qubits. This drops the lower bound of the qudit model to $d > 0$. As a result, we introduce the new term Qusym (for “quantum symbol”) to describe this specialised qudit variant. When formulated as a density matrix, the qusym model therefore precisely maps onto Shannon’s notion of probabilistic selection from a finite vocabulary of unique symbols. That is, if each of the basis vectors defining a qusym denotes a unique symbol, encoding must satisfy the condition that the total entropy available to that qusym sums to 1. Therefore where $\rho$ denotes the density matrix and $\rho$ individual selection probabilities (for the pure states $|\alpha_d\rangle$):

\begin{eqnarray}
    \rho \equiv \sum_{0}^{d-1} p_d|\alpha_d\rangle\langle\alpha_d|
\end{eqnarray}

and
\begin{eqnarray*}
    tr(\rho) = 1    
\end{eqnarray*}
Which maps to the total available von Neumann entropy, as in:

\begin{eqnarray}
    S = \sum_{0}^{d-1} p_d \ln p_d 
    = -tr(\rho \log)
\end{eqnarray}

The rest of this paper focuses on the use of the qusym model and restricted/partial tensor product use (entanglement) in a framework similar to that postulated in VSA.

\section{Monogomously Entangled Quantum Systems as Semantic Triples}
\label{semantic_triples}
Given the prior application of standard approaches to quantum computing for modelling formal semantics \cite{widdows20} \cite{smolenksy90} \cite{kanerva09} \cite{clark07} \cite{widdows08} \cite{mitchell10} \cite{baroni10} \cite{turney12} \cite{garg19} \cite{cohen12} \cite{cohen11}, it becomes possible to consider combining the principles of distributional representation \cite{luhn59} \cite{jones97} \cite{maron60} \cite{rijsbergen04} \cite{widdows04} \cite{mikolov13} \cite{pennington14}, with the properties of quantum entanglement \cite{einstein35} \cite{schrodinger35} \cite{bennett2000} \cite{bennett98}, to model predicate relationships between two or more quantum systems in a finite vector space \cite{widdows20} \cite{cohen12} \cite{cohen15}. This creates \textit{quantum triple} arrangements and hence sub-graph structures which explicitly provide the “correlation” between “physical or conceptual entities” originally shunned by Claude Shannon when he dismissed meaning in his seminal 1948 paper on Information Theory \cite{shannon48}. 

%-------------------------------------------
\begin{figure}[H]
\centering
\includegraphics[width=\columnwidth]{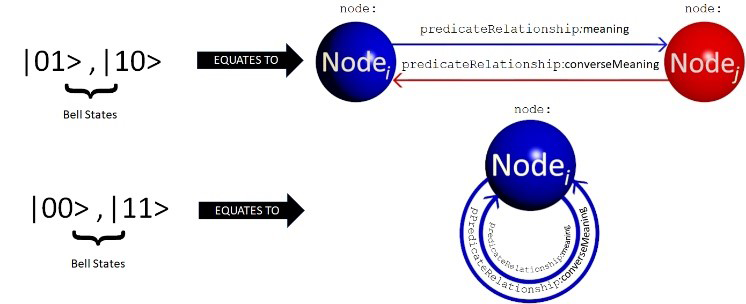}
\caption{Triple Patterns to Bell States Mapping}
\label{Figure 3}
\end{figure}
%-------------------------------------------

Quantum entanglement is a phenomenon that sees two or more quantum systems interact and share a relationship so that the quantum state of one cannot be described independently of the others. It is therefore usual to consider two maximally entangled qubits configured to correspond to one of the four possible Bell states \cite{nielsen2000} – although entanglement need not be maximal in all perceivable cases. This assumes that both systems contribute equally to their joint entangled state, although for all practical intents and purposes it is important to remember that these are physical systems and so the exact nature of these contributions is more variable. 

In this formation, as under standard (binary) quantum computing conditions, the spin (dipole magnetic moment) of the entangled systems involved must polarise to align with pure ground states of “up” or “down”, along some measurement basis. These alignments correspond to the binary values of zero or one, representing the computational basis states, and, not incidentally, this complies with the standard approach for calculating information entropy using $log_{2}$ . This is therefore sufficient to articulate the Bell states, which can then be mapped onto triple metapatterns, as shown in Figure \ref{Figure 3}: 

In providing this mapping, it should be noted that the binary encoding system normally adopted in quantum computing applies merely as a consequence of a reduction down to the simplest form of freedom of choice \cite{shannon48} and the associated convenience of Boolean logic for switching in digital computing. That is, meaning associated with the strings that are the symbols “0” and “1” maps to familiar primitives like negative or positive states, “no” or “yes” and the switching between two constant voltages in an electrical circuit. This binary system adds little when wanting to encode information, which can be far richer and more complex than binary primitives naturally cater for. In such cases, it might be preferable to consider an encoding range of, say, 128 when modelling information encoded using standard American Standard Code for Information Interchange (ASCII) format. Such encoding therefore implies the use of $log_{128}$ entropy calculation for single symbol strings, given that each such symbol is drawn from a vocabulary of 128 similarly encoded unique symbols.

%-------------------------------------------
\begin{figure}[H]
\centering
\includegraphics{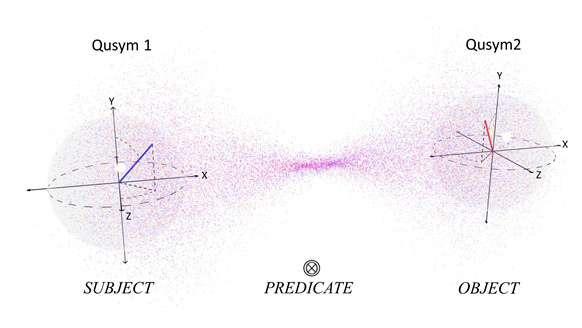}
\caption{An Illustration of Quantum Entanglement as a Simple Triple}
\label{Figure 4}
\end{figure}
%-------------------------------------------

As conceptualised in Figure 4, the use of monogamously entangled quantum systems to act as paired semantic corollas is interesting. Specifically, any such singularly involved physical system $n^{quantum}$ can be configured to encode the semantics of some entity, concept, state or string drawn from a finite set of allowable values $N^{quantum}$. Likewise, non-directional predicate relations $p^{quantum}$can be configured from a finite set of allowable values $P^{quantum}$, as can directed predicates, $p_{directed}^{quantum}$ from the finite set $P_{directed}^{quantum}$. The latter can then be used to set the degree of entanglement contributed by any semantically configured quantum system when entangling with another quantum system. That is, both the vector state of that quantum system and its degree of entanglement with a second quantum system must be set for the legitimate structure of a corolla to form. It also mandates that corolla cannot be physically instantiated in isolation due to their inherently entangled nature. All corolla must therefore be part of an entangled ordered pair of corollas $(c_i,c_j )$  , as stated previously, which together denote the single triple $t_{i,j}$.

In isolation, and assuming that \texttt{person}:{\bf Bob} represents a male, every valid combination of $\lim \atop
{_N^{quantum},P_{directed}^{quantum}}$ (${n^{quantum},p_{directed}^{quantum}}$), generates a pure state physical corolla, where, for instance, the possible pure states of the corolla (\texttt{person}:{\bf Bob}, $p_{directed}^{quantum}$) would be (\texttt{person}:{\bf Bob}, \texttt{kin}:{\bf ParentOf}) and (\texttt{person}:{\bf Bob}, \texttt{kin}:{\bf HusbandOf}), if  $P_{directed}^{quantum}$ is configured to represent the limiting set of semantic relationships in \{\texttt{kin}:{\bf ParentOf}, \texttt{kin}:{\bf ChildOf}, \texttt{kin}:{\bf HusbandOf}, \texttt{kin}:{\bf WifeOf}\}. In highlighting this restriction, it is important to note that this set is itself made up of converse relationship-pair subsets, as in \{\texttt{kin}:{\bf ParentOf}, \texttt{kin}:{\bf ChildOf}\} and \{\texttt{kin}:{\bf HusbandOf}, \texttt{kin}:{\bf WifeOf}\}, covering both sides of all allowed entanglement relationships formed between the two quantum systems involved. As such, not only do combined corolla individually contribute opposite meanings, but the amount of physical entanglement they contribute can be thought of as having a form of direction, so that one corolla contributes positively to its entanglement relationship and the other negatively and in equal measure. In this way, the overall directional entanglement contribution cancels itself out, but the modulus of both equals the total degree of entanglement between the two contributing quantum systems. This is the same as saying that both systems contribute half the amount of entanglement entropy involved in their jointly entangled state. 

From a quantum configuration standpoint then, the corolla (\texttt{person}:{\bf Bob}, \(p_{directed}^{quantum})\) corresponds to any classically quantum system, \(n^{quantum}\), that is \texttt{person}:{\bf Bob} via a quantum qusym (not a qubit or qudit) whose measurement yields any of d possible pure states, where:

\hfill \break
\setlength{\leftskip}{0.5cm}
\begin{enumerate}[(a)] % (a), (b), (c), ...
\item $n^{quantum}$ is a quantum system (qusym) used to represent the semantics of an entity, concept, state or string. The condition of any such system is measured as a pure state unit quantum vector corresponding to one of a set of allowable vectors $N^{quantum}$.
\\ \\
\item $N^{quantum}$ is the set of all allowable pure states for the quantum system $n^{quantum}$. That is a set of all allowable vectors corresponding to all available entities, concepts or strings. It is equivalent to a vocabulary containing relevant/allowable entities, concepts or strings as explicitly conceived by Shannon \cite{shannon48}.
\\ \\
\item $d_{node}^{quantum}$ is the finite cardinality of the set $N^{quantum}$. That is: $d_{node}^{quantum} = |N^{quantum}| \in \mathbb{N}_1$
\\ \\
\item $p^{quantum}$ is the degree of entanglement (entanglement entropy) between two quantum systems (qusyms) used to represent a non-directed predicate relationship between the two. This is a scalar value representing the degree of entanglement between two quantum systems.
\\ \\
\item $P^{quantum}$ is the set of all allowable non-directional quantum entanglement scalars. It is equivalent to a vocabulary containing all relevant/allowable non-directed predicate relationships.\newline\newline The degree of entanglement (entanglement entropy) between two quantum systems (qusyms) used to represent a directed predicate relationship between the two quantum systems. This is a scalar value representing the degree of entanglement contributed by either of two monogamously entangled quantum systems, in either:
\begin{eqnarray}
    + {p^{quantum} \over 2} \in \mathbb{R}_0^+   
\end{eqnarray}
or 
\begin{eqnarray}
    - {p^{quantum} \over 2} \in \mathbb{R}_0^-      
\end{eqnarray}
\\ \\
\item $P_{directed}^{quantum}$ is also
\begin{eqnarray}
    \left( 2|\xrightarrow[p_{directed}^{quantum}]{}| \right) \in P^{quantum} \in \mathbb{N}_1
    \\
    \left( 2|\xleftarrow[p_{directed}^{quantum}]{}| \right) \in P^{quantum} \in \mathbb{N}_1
    \\
    |\xrightarrow[p_{directed}^{quantum}]{}|+|\xleftarrow[p_{directed}^{quantum}]{}| \in P^{quantum} \in \mathbb{N}_1
    \\
    \xrightarrow[p_{directed}^{quantum}]{}+\xleftarrow[p_{directed}^{quantum}]{} \equiv 0
\end{eqnarray}

The above mandates that any grounding of $\xrightarrow[p_{directed}^{quantum}]{}$ must always be the negative equivalent of $\xleftarrow[p_{directed}^{quantum}]{}$ so that if, for instance, $\xrightarrow[p_{directed}^{quantum}]{}=0.2$, then $\xleftarrow[p_{directed}^{quantum}]{}=-0.2$. It also states that in this example,  the total amount of entanglement entropy between the entangled quantum systems involved must be $0.4$.
\\ \\
\item $P_{directed}^{quantum}$ is the set of all available quantum entanglement scalars.
\\ \\
\item $d_{directed}^{quantum}$ is the cardinality of the set $P_{directed}^{quantum}$, that is $d_{directed}^{quantum} = |P_{directed}^{quantum}| \in \mathbb{R}_1$
\\
\\
\end{enumerate}
\setlength{\leftskip}{0.5cm}
Note also that:
\begin{eqnarray}
    \left|\xrightarrow[p_{directed(i)}^{quantum}]{}\right|\oplus\left|\xleftarrow[p_{directed(i)}^{quantum}]{}\right| \triangleq p^{quantum(i,j)}
\end{eqnarray}
The combination of the directed predicate pair between two quantum nodes, $n^{quantum(i)}$ and $n^{quantum(j)}$ is exactly equivalent to the the non-directed entangled relationship, $p^{quantum(i,j)}$, between them.
\\\newline
Therefore, $p^{quantum(i,j)}$ is an intrinsic attribute \cite{deutsch15} of the combined system $n^{quantum(i)} \oplus n^{quantum(j)}$ which is $t_{i,j}$ but is not another intrinsic attribute of either $c_i$ or $c_j$.
\\\newline
As such, quantum corollas can be represented via a unit vector, scalar pair, as in: 

\hfill \break
\setlength{\leftskip}{1.0cm}
$(n^{quantum}, p_{directed}^{quantum})$ is a vector that is a scale pair representing a quantum corolla, without explicitly considering the direction of any contribution made towards the overall degree of entanglement.
\\ \\
$(n^{quantum}, p_{directed}^{quantum})$ is a subject (left-hand) corolla in a quantum triple pair, which is equivalent to:
\begin{eqnarray}
    n^{quantum} \oplus \xrightarrow[p_{directed}^{quantum}]{}
\end{eqnarray}
\\ \\
$(n^{quantum}, p_{directed}^{quantum})$ is an object (right-hand) corolla in a quantum triple pair, which is equivalent to:
\begin{eqnarray}
    n^{quantum} \oplus \xleftarrow[p_{directed}^{quantum}]{}
\end{eqnarray} \newline Where $n^{quantum}$ is a unit scale vector $|\psi_i \rangle$ in the state space $\mathbb{C}^{d_{node}^{quantum}}$ and $p_{directed}^{quantum}$ some scalar real value representing one half of the degree of entanglement associated with $n^{quantum}$ when monogamously entangled with another quantum system (qusym). Therefore, without loss of generality, all corolla can be physically denoted as:
\begin{eqnarray}
    c_i = \left( n_{node}^{quantum},p_{directed}^{quantum}\right)
\end{eqnarray}
\hfill \break
\setlength{\leftskip}{0.5cm}
Likewise, a quantum semantic triple can be denoted as a composite pair of quantum corolla:
\begin{eqnarray}
    t_{i,j} = (c_i,c_j) \triangleq |\psi_{i,j} \rangle = \sum_{ij} k_{i,j}|\psi(i)\rangle \otimes |\psi(j) \rangle
\end{eqnarray} 
\hfill \break
\setlength{\leftskip}{1.0cm}\newline Where: $k_{ij}$ is the complex coefficient corresponding to the state $(i,j)$ such that $\overline{k_{ij}}k_{ij} \geq 0$ is equal to the probability of quantum semantic triple $t_{ij}$ collapsing to the pure state (${i,j}$). Here, $\overline{k_{ij}}$ denotes the complex conjugate of $k_{ij}$ and each of $i$ and $j$ equals any value combination of allowable ground states for $n^{quantum}$ nodes and $p^{quantum}$ directed predicate relationships.

\hfill \break
\setlength{\leftskip}{0.5cm}\newline Also:
\begin{eqnarray}
    t_{i,j} = (c_i,c_j) \cong \lim_{ \xrightarrow[p_{directed(i)}^{quantum}]{},\xleftarrow[p_{directed(i)}^{quantum}]{}}|\psi_{i,j}\rangle= \sum_{ij} k_{i,j}|\psi(i)\rangle \otimes |\psi(j)\rangle
\end{eqnarray} \newline i.e. the specialised outer product produced by the partial entanglement of $n^{quantum(i)}$ and $n^{quantum(j)}$ is restricted to only cover the relationship created when $c_i$ and $c_j$ bind together through entanglement.
\\ \\
\hfill \break
\setlength{\leftskip}{0.5cm}\newline As an example, the semantics of the parent-child relationship between \texttt{person}:{\bf Bob} and \texttt{person}:{\bf Alice} physically grounds to a monogamously entangled pair of quantum systems (qusyms) as:
\\
\begin{eqnarray}
    c_{Bob:parent}= \left(n_{node(person:Bob)}^{quantum}, p_{directed(kin:ParentOf)}^{quantum}\right)
\end{eqnarray}
\\
\setlength{\leftskip}{1.0cm}\newline i.e. $n_{node(person:Bob)}^{quantum}$ is grounded to \texttt{person}:{\bf Bob} and $p_{directed(kin:ParentOf)}^{quantum}$ is ground to \texttt{kin}:{\bf ParentOf}.
\\
\begin{eqnarray}
    c_{person:Alice:kin:children}=\left(n_{node(person:Alice)}^{quantum},p_{directed(kin:children)}^{quantum} \right)
\end{eqnarray}
\\
i.e. $n_{node(person:Alice)}^{quantum}$ is grounded to \texttt{person}:{\bf Alice} and $p_{directed(children)}^{quantum}$ is ground to \texttt{kin}:{\b ChildOf}.

\begin{eqnarray}
    t_{person:Bob,person:Alice} = \left( c_{person:Bob},c_{person:Alice}\right) 
    \triangleq | \psi_{person:Bob,person:Alice}\rangle
\end{eqnarray}

\begin{eqnarray}
= { \sum_{v} { k_{Bob,Alice} |\psi\left((person:Bob):(kin:parentOf)\right)\rangle\otimes|\psi\left((person:Alice):(kin:childOf)\right)\rangle }  } 
\end{eqnarray}
Where $v = (person:Bob):(parentOf),(person:Alice):(kin:childOf)$.

\setlength{\leftskip}{0cm}

\section{Further Research}
\label{further_research}
Given that the idea of corollas embodies notions of both \textit{self} (node) and \textit{association} (edge) at axiomatic levels, it naturally complements ongoing research into areas like information theory, information compression, encryption and formally symbolic grammars. This especially includes initiatives aimed at representing natural language and the specification, design and build of general data repositories on quantum hardware.

\section{Conclusions}
\label{conclusions}
For reasons of convenience, the complications brought about by vector duality and axiomatic interval definition are often overlooked. However, in the realms of information theory and formal semantics, they remain extremely important. The descriptions associated with both the whole and its parts contribute materially to any complete and proper notion of meaning. That is especially true of the associations introduced through the axiomatic closure of primitives and any associative composition; so that any complete attempt to assess the semantics of a multi-part system must account for the relationships involved between all sub-components and at all levels of abstraction. QC analysis techniques targeting the pure-state systems resulting from linear operations find this difficult to accommodate, highlighting a weakness in contemporary practice. This is seen, for instance, when mixed state analysis techniques, like density matrix inspection, are considered to analyse the axiomatic properties of pure state multi-part systems like those found in quantum entanglement. 
Furthermore, convention in IR, Quantum Information Theory (QIT) and QC tends to treat associative composition as an extraneous concern, implying that work with vectors spaces requires practitioners to have significant prior knowledge of the relationships between any systems, concepts or states being investigated. This also highlights a general weakness in practice. Conversely, several studies have explored the use of tensor product arrangements to model natural language and the information content embodied within. These are founded on original work to apply the properties of \textit{entanglement} (via specialized linear product calculation) to logically bind language primitives together \cite{smolenksy90} \cite{kanerva09} and have since been extended to model grammatical and semantic composition \cite{clark07} \cite{widdows08} \cite{mitchell10} \cite{baroni10} \cite{turney12}, general phraseology \cite{clark07} and knowledge representation \cite{cohen12} \cite{cohen15}. However, these do not treat associative relation as an assignable primitive - other than the binary ability to be bound or not - and so ignore the value of axiomatisation. More importantly, they miss any potential for associative (binding) relationships to carry useful semantics.

By introducing the notion of corollas, informational structure becomes explicitly embedded in any vector space under consideration. This allows semantics to be expressed across both the nodes and edges present in the graph that is that structure, thereby taking novel advantage of the variability associated with the bilinear tensor product \textit{binding} between node primitives in composite pure state systems like entangled node pairs. This lowers the level of abstraction used for semantic representation seen in previous studies \cite{cohen12} \cite{cohen15} so that it becomes equally axiomatic across both nodes and their predicating relationships. It also doubles the range of options available for axiomatic semantic expression. Consequently, the only extraneous concern for implementation becomes that of string vocabularies acting as pointers into the vector space. This extends the reach of both Boolean and quantum logic across vector spaces from first principles. Whereas the quantum extends classical Boolean logic to include capabilities like interpolation and extrapolation \cite{widdows20}, injecting semantic structure via corollas strengthens the findings of previous research \cite{cohen15} \cite{cohen11} and extends quantum reasoning to include graph-theoretic opportunities for inference and formal deduction. Both are important to emerging fields like Quantum AI \cite{widdows20} \cite{cohen12} \cite{cohen11} and Logical Neural Networks \cite{riegel20}.

\section{Appendix A: History}
\label{appendix_a}
When Claude Shannon extended the work of Nyquist and Hartley \cite{nyquist24} \cite{hartley28} \cite{goodman17} and published his foundational paper on Information Theory in 1948 \cite{shannon48}, it proved a tipping point for technological advance. Its title is somewhat misleading though, in that it concentrates on the fundamental limits of signal processing and the physics of communication \cite{deutsch15} rather than any definition of information per se. Shannon himself openly acknowledged this fact and was troubled by the philosophical interpretations of information attached to his work; as is well documented in his approach to von Neumann for advice on how to label his ideas \cite{goodman18}. But by far his best-known concern lay with the implications of meaning within the definition of information itself. His paper even calls this out and deliberately plays down the role of meaning in the face of more pressing challenges: 
\begin{quote}
    “The fundamental problem of communication is that of reproducing at one point either exactly or approximately a message selected at another point. Frequently the messages have meaning; that is they refer to or are correlated according to some system with certain physical or conceptual entities. These semantic aspects of communication are irrelevant to the engineering problem. The significant aspect is that the actual message is one selected from a set of possible messages. The system must be designed to operate for each possible selection, not just the one which will actually be chosen since this is unknown at the time of design.”\cite{shannon48}
\end{quote}
\

With this, Shannon forced a remarkably practical stance. To him at least, meaning was merely an obstacle in the way of optimisation, so he expelled it from his theory \cite{shannon48}. Yet the very purpose of communication is to share knowledge and ideas, to trade in meaning in order to become more widely understood. More specifically, meaning underpins how communicating parties define the symbols they wish to share, along with the operations that might be applied over them. The convention of Computer Science refers to this construction, made from symbols via said operations, as \textit{Strings} and from that foundation, one can begin to understand why the ideas of communication and meaning must be included in any complete theory of information.

\section{Appendix B: Entropy and Time}
\label{appendix_b}
At its heart, Shannon’s Information Theory is one of entropy and deals with levels of uncertainty when sending and receiving messages made of strings. Shannon Entropy is therefore probabilistic and, by implication so is the abstract communication process laid out in his 1948 paper. In simple terms, it is about accurately transforming information from one format into another then back again. In that regard, it is inherently \textit{computational} but not necessarily \textit{informational} – viz. functions that contain operations map a domain of strings to a co-domain of other strings (computation) and back again, but this does not necessarily imply purpose or utility within any domain of interest (information). To understand why, it is possible to consider two questions by way of a simple thought experiment. First: \begin{quote}
    How much computation would be left in the world if it were possible to stop time? 
\end{quote}

The answer is none, since computation is a form of \textit{action} targeted at \textit{change}, which by definition entails work done over \textit{time}. This is even true of “no-op” type instructions, where the \textit{representation} \textit{state} of a machine remains unchanged but its \textit{physical} \textit{state} transforms from one immediately before the no-op execution, to one immediately after. More broadly, time is the measure of change in the world and, according to convention, without it computation misses an ingredient that is both necessary and sufficient in its definition. Simply, computation is a transformational act that demands time’s presence. 

Next, ask a similar question:
\begin{quote}
    How much information would be left under the same circumstance?     
\end{quote}

The answer is all of it. Information exists independently of any computation that might operate over it and therefore is not reliant upon time. This leads to an alternative interpretation of Shannon Entropy which excludes time and is not probabilistic. In this formulation, information is defined purely as a measure of distinguishability \cite{deutsch15} between individual strings and their background context - in the form of a source string pool or vocabulary. This implies that any such vocabulary must contain at least two such strings and that both are encoded using the exact same protocol, to ensure that measurement can reliably and accurately determine them apart \cite{deutsch15}. In this process, there is no notion of time. Strings just come into being through the act of measurement, regardless of whether preparation is needed or not. There is no “before” or “after” \cite{deutsch15}.

Even with this strengthening in place, however, information theories in general remain lacking, not least because of this very emphasis on distinguishability through difference. 

As a basis for the definition of information, difference alone is problematic. Without contrast to introduce punctuation, the resulting continuum would be closed to all but assessment as a singular smooth whole. In other words, for a unit of information to form, there is a need to describe both:
\begin{enumerate}[(a)] % (a), (b), (c), ...
    \item The \textit{interval(s) of equivalence }present across its useful characteristics; and 
    \item The boundaries that limit that equivalence and constrain its usefulness.
\end{enumerate}

Classical information theory makes this clear by separating strings within vocabularies, thereby wrapping informational content into neatly discrete micro-parcels, or quanta, that are its units for transmission. More formerly, information is encoded as strings whose meaning is embodied via the symbols as a proper subset of all of those available within a domain of interest, and which are constructed using the operations under which the total symbol set is closed. This introduces the idea that strings conform to a \textit{grammar} that “wraps” both the valid symbols and their operations into a single structure. A string may therefore be said to conform to a given grammar $CG$, which itself contains the set of valid symbols ($SY$) and the set of valid operations ($OP$).
\setcounter{equation}{22}
\begin{eqnarray}
    s: \forall s \in CG = SY\{sy_0,sy_1,...,sy_n\},OP\{op_0,op_1,...,op_n\}
\end{eqnarray}

More elegantly, any set of strings S can be considered as closed under the grammar CG when the latter provides an organised whole of valid symbols and corresponding operations. 
\begin{eqnarray}
    S^{CG}
\end{eqnarray}

Information theory adds little here however, as it does not explain how symbol or string separation should take place, or provide any formulaic understanding related to equivalence or similarity. That is to say, classical information theory does not address the fundamental challenge of symmetry breaking essential to information unitisation – a property which is vital to both information encoding and measurement. This highlights the role of symmetry and asymmetry within the definition of information - or in more familiar terms, the importance of difference and equivalence in understanding what information \textit{is}.

The credibility of classical information theory is not in question. Nevertheless, the arrival of Quantum Information Theory (QIT) \cite{bennett73} \cite{bennett98}, the earlier advent of  Landauer’s principle \cite{rolf61} and the more recent appearance of Constructor Theory \cite{deutsch15} have openly challenged its completeness. QIT in particular seeks to extend and complete classical theories “…somewhat as complex numbers extend and complete the reals” \cite{bennett98}. But, even today, the limits of QIT remain unclear especially given that it “never gets round to specifying what exactly it is referring to as ‘quantum information’ nor its relation to classical information.” \cite{deutsch15}. So, QIT “is not, despite the name, a theory of a new type of information, but only a collection of quantum phenomena that violate the laws of classical information”. This makes it clear that further extensions are possible, perhaps to include principles that are somehow deeper and applicable beyond both quantum theory and Shannon’s ideas \cite{deutsch15}. For instance, although QIT and quantum computing more generally provide some insight into symmetry breaking for information unitisation, they still have little to say about the thorny subject of meaning. In contrast, the crossover between formal semantics, graph theory and quantum mechanics has been studied in detail \cite{w3c13} \cite{rijsbergen04} \cite{widdows04} \cite{masterman05} \cite{charniak76} \cite{zipf49} \cite{rijsbergen95}, seeing many practical advances in areas like Information Retrieval (IR) \cite{rijsbergen04} \cite{sihare16} \cite{landauer98} \cite{luhn59} \cite{jones97} \cite{maron60} \cite{luhn58} \cite{jones73} \cite{masterman05}. For that reason, this work is sometimes loosely labelled as IR for convenience, but it extends into related fields like Knowledge Representation (KR). KR therefore represents facts using strings and applies logical reasoning to infer new facts and make deductions - in common with the set theory ideas introduced above. Examples of symbolic and logical frameworks include First Order Logic (FOL), the Resource Description Framework (RDF) \cite{rdfcore04}, the Web Ontology Language (OWL) \cite{owl08}, Semantic Web Rule Language (SWRL) \cite{swrl04} and the Frame Language \cite{garg19}.

Given the clear links between QIT and IR/KR then, it is perhaps surprising that there has been little overlap between IR/KR research and QIT – in particular on how quantum computing might directly incorporate semiotic data structures. Regardless, deep links between symmetry, entropy and meaning are now understood \cite{shukun96} \cite{symm05} \cite{penrose04} \cite{linden09} \cite{riemann08}, especially given symmetry’s central role in formal semantics \cite{wittgenstein33} \cite{chierchia85} \cite{cyganiak14} under the guise of association through shared trait or predicate. Consequently, just as a complete theory of information must extend down to the substrate that is quantum mechanics, the same must be true for any extension that covers semantics and knowledge representation. 

To understand how practice might achieve such representation, it is first important to recognise that entropy is not just concerned with randomness, irreversibility, or disorder. In its most traditional form it quantifies the ability to do \textit{work} in thermodynamic terms \cite{muller07}, and so is deeply linked with ideas on computability. But at face value that presents a challenge, as entropy is a scalar quantity and so lacks the directional element needed to describe structural arrangements like those found in computing. Time provides a proxy for this in bounded thermodynamic systems, in so much as it describes the entailment between adjacent entropic states. That is to say, the \textit{transition} between differing levels of countable entropy strictly follows the structure that is the arrow of time.  With time removed, the need for structure can be overcome by superimposing graph-based models like those found in state transition diagrams \cite{kauffman2000} \cite{kauffman08} \cite{kauffman08_1}. This therefore implies that vector-based approaches can model many entropic systems. When interpreted liberally, this expands to cover a broad range of graph-theoretic models, including spin networks, and in such cases measures of entropy can be used as metadata for classification and identification purposes. 

\section{Appendix C: Conservation Laws, Formal Semantics and Triple-based Data}
\label{appendix_c}
When graph-theoretic techniques are used to study non-ergodic systems, it is common to see an overall drop in entropy, a corresponding decline in the ability to undertake \textit{work} freely and an increase in association levels between the subsystems or micro-states involved \cite{kauffman08_1}. The latter is notable because it demands the sharing or conservation of at least one property across two or more constituent parts or states, thereby adding symmetry to regions of any parent system or macro-state. The first law of thermodynamics accommodates this through Noether’s Theorem and its emphasis on conservation laws over differentiable symmetries \cite{davies07}, with notable examples being seen in chemical bonds, social or commercial interaction and so on. In all such cases, graph associations allow ontological analysis through graph structure, with related metadata contributing towards semantics. Accordingly, such graphs can represent Semantic Networks or Knowledge Graphs \cite{lehmann92} \cite{simmons63} \cite{simmons82} constructed from basic building blocks known as triples \cite{guha92}. These correspond to the connection, conservation, relationship or symmetry between two distinct graph nodes as unary axioms about some shared unary characteristic, trait, state or theme. This implies bidirectional unary entailment and provides a rudimentary mechanism for combining difference and similarity. Graph nodes and edges can therefore act as logically axiomatic, unary strings, concepts, classes, clauses, states or concepts \cite{garg19}. Furthermore, graphs can be formalised as mathematical objects using the same principles of set theory introduced in the preceding sections – viz. a graph $G$ comprises a set of vertices (nodes) $N(G)$ and a set of edges $E(G)$ that describe its constituent elements and linkages, and therefore its structure. In terms of information theory, what each node \textit{is} or represents depends on the domain of discourse and so also defines the nature of the edges that link to it. Just as a grammar bounds what a string can \textit{be}, so there are corresponding criteria applied to graphs.

The idea of axiomatic unary isolation is important here; that individual nodes and edges can only ever act as property holders specific to their own unique existence. They are therefore units of self-evident truth that require no further proof, and in that regard, are quite separate from any implied wider contexts. Just as there is no “before” or “after” in information-theoretic terms, there is no “other” than that which is spelt out logically by the triples they create and their contribution to the parent graph that is their world. Without this possession, connection and containment of truth, that world would, for all practical intents and purposes, be useless \cite{kauffman2000}. That is, any physical realisation of the information contained could not interact and exchange energy in order to do beneficial work under the thermodynamic definition of entropy \cite{west171}. 

\section{Appendix D: Infinite Measurement Schemes, Turing Completeness and String Uniqueness}
\label{appendix_d}
To further expand on the shortcomings of current information theories, it is important to restate the dominance of simplified freedom of choice \cite{shannon48} in contemporary information technology. Or as Shannon put it:
\begin{quote}
    “…what would be the source you might have, or the simplest thing you are trying to send. And I’d think of tossing a coin.” 
\end{quote}

Consequently, quantum computing’s favoured polarising measurement approach, in Stern-Gerlach \cite{gerlach22}, falls in line and defaults to binary operation. This therefore “snaps” the magnetic moment of a quantum system’s spin into one of two alignments along some measurement basis as quantum coherence occurs and thereby provides a convenient means to “map” measurements from the continuousness of Hilbert space down onto the discrete notation of the binary number system. Binary alignment is not some innate property of any quantum system though. So, without the practicalities of coherence brought about by measurement, quantum information defaults to a continuous vector space about some arbitrary spin moment orientation. This reminds that the quantum systems contained within and manipulated by quantum computers are not predisposed to bi-state polarisation under all theoretically plausible measurement schemes. 

In a similar vein, quantum entanglement is also not predisposed to bi-state polarisation; as in “entangled” or not. It is simply a measure of the degree of quantum entanglement in a multi-body quantum state and is defined as the Von Neumann entropy of the reduced density matrix of a given quantum subsystem, traced over other subsystems \cite{entr20}. It can be expressed for any multi-body quantum system using the singular values of a Schmidt decomposition of state vectors. As any pure quantum state can be written as $|\psi\rangle=\sum \alpha_i|\mu_i\rangle A \otimes |v_i\rangle B$ where $\mu_i\rangle A$ and $|v_i\rangle B$ are orthonormal states in subsystem $A$ and subsystem $B$ respectively, the entropy of entanglement is simply the scalar:
\begin{eqnarray}
    - \sum_{i=1}^{i=2^2} |\alpha_i|^2\log(|a_i|^2)
\end{eqnarray}
This can be generalised to:
\begin{eqnarray}
    - \sum_{i=1}^{i=d^2} |\alpha_i|^2\log(|a_i|^2)
\end{eqnarray}
for encoding variable base encoding systems of base $d$. This gives the degree of association between entangled quantum subsystems and is analogous to notions of symmetry, conservation and equivalent trait. It also explains that entanglement entropy can encode information and therefore implement graph-based denotational semantics \cite{cohen12} \cite{cohen15}. Furthermore, in situations where the entangled systems in question have been precast in some explicit frame of reference, perhaps through their position \cite{kauffman93} within some parent graph, information entropy offers the potential to implement an explicit framework for formal semantic ontologies in quantum information. 

In purely theoretical terms then, it is reasonable to consider tessellating (a) Hilbert space into more than two subspaces for the purpose of spin measurement - much like dividing a clock face into minutes, seconds and so on, where the measurement of time is itself in this way a superposition of three number bases comprising hours (base 12, 24), minutes (base 60) and seconds (base 60). Therein, the greater the degree of division the more expressive any associated encoding system will be. Likewise, the greater the range of strings open to reference and the more compressed any information encoded within the quantised subspace that is the tessellated metric system. This is equivalent to using the numeric base open to any quantum system, $d$, to take logarithms in conventional formulations of information entropy, so that 2 tessels (primitive measurement intervals) equate to the standard $log_{2}$ formulation, 128 tessels gives $log_{128}$, and so on, out to $log_n$. Thus, as where in standard implementations of information theory, the total information entropy in a string scales \textit{horizontally} when its encoding base remains constant but length increases, total entropy scales \textit{vertically} when a string’s length remains constant but its associated encoding base increases - in agreement with Nyquist’s findings and practical application in areas like the fractal compression of image data \cite{fisher94}. By analogy, this may be considered a matter of resolution in which the vertical scaling does not change the string’s image but the prevision and sharpness with which one can model that image. The standard horizontal Shannon-scaling of entropy therefore produces an exponential increase in information, whereas scaling vertically results in power-law growth, as seen, for example, in biological systems of varying size \cite{west171}.

This leads to an interesting observation. Through its mathematical continuity, it is theoretically possible to consider subdividing Hilbert space infinitely, thereby creating an infinite pool of strings for information encoding and hence presenting the opportunity for infinite semantic expression. This is equivalent to the “one hot encoding” method for mapping strings to independent base vectors \cite{widdows20} \cite{geron19} and is a product of Hilbert spaces themselves being defined over certain domains (abstract geometries with respect to strings) and closed under certain operations (in this case, division). Likewise, this also offers the opportunity for infinite information compression, since the constraints of scalar information loss that arise from Shannon’s horizontal-only measure of entropy are compensated for in the vertical. Analogously, it is worse than thinking of the Bloch sphere as a “golf ball” print head containing multiple infinities of unique printable characters, given that tessels, as base vector endpoints, can also potentially be used to introduce polysemy of meaning through implied superposition \cite{geron19} and therefore undermine the need for universal axiomatisation in information representation. 

In turn, this raises a dilemma, in that infinite division would actually impede practical application. 

For all that semantic scholars often play down their importance \cite{wilks08}, four basic language types result from frameworks like Chomsky's hierarchy of grammars \cite{chomsky59}. These are Regular, Context-Free, Decidable and Turing-Recognisable, and nest so that the right-most encapsulates those on the left, and in contemporary computer science this hierarchy builds on the foundational teachings of Chomsky, Turing and Shannon himself. Chomsky’s hierarchy therefore applies to a broad range of formal grammars and is a common linguistic tool. Even so, this framework is not applicable to all forms of language \cite{wilks19} \cite{jager12}, so, when developing a view of information supported by an infinitely divisible vector space over the complex numbers, an information model is formed that even Turing-recognisable grammars cannot accommodate. With such languages, the general aim is to be provably decidable and in that they specifically reject strings not in their vocabulary, in the desire to work towards some halt condition - this also implies the rejection of “symbols”, “vocabulary entries” or “qualified semantics”, or in this case vector pointers to all of the above. Or in generalised linguistic and systems terms, they support the need for statements to be meaningfully closable and therefore practically \textit{useful} \cite{west171}. This builds on the well-established base principles of decidability and halting in the fields of computer science and discrete mathematics.

Without the ability to limit the expression available to any formally describable language, the need to halt and be \textit{useful} becomes an insurmountable problem, simply because, for instance, it could take an infinite amount of time to look up valid strings. In practice, however, this problem can potentially be overcome in fields like quantum computing by analysing the error density matrices produced by standard Stern-Gerlach-like measurement and rounding to intervals corresponding to the intra-quanta dimensions of any assigned metric system. In that, the summarised outputs translate into a finite set of standardised vector intervals pointing to valid strings in any assigned vocabulary. For all intents and practical purposes, this therefore makes semantics an informational construct - as a function of quantised entropy and a by-product of measurement stability and accuracy. It also profoundly supports the basic premise of QIT, in that either information is a consequence of quantum mechanics or quantum mechanics is a consequence of information.

\section{Appendix E: Shannon's Communication Process and a Semantic Information Theory}
\label{appendix_e}
When Shannon’s model for communication, as shown in Figure 5, is considered in detail it becomes clear that the inclusion of semantics has always been there, in that both transmitter and receiver are expected to implicitly “understand” how to interpret sequentially transmitted strings at either end of a communication channel (defining the need for mutually understood communications protocols). Specifically, the rules governing translation in and out of these protocols are hard-wired so that any semantics associated with information transmission are embedded into the communication process itself. This makes Shannon’s model context specific from its first principles and as such, also associates it with one of Chomsky’s innermost grammars due to the forced implication that only finite-length vocabularies can be used. This creates the high degree of control Shannon deliberately sought to reduce redundancy in information transfer, but in that there is no attempt to “explain” the information being sent. That is left to processes extraneous to the mechanics of message passing. 
%-------------------------------------------
\begin{figure}
\centering
\includegraphics[width=0.8\columnwidth]{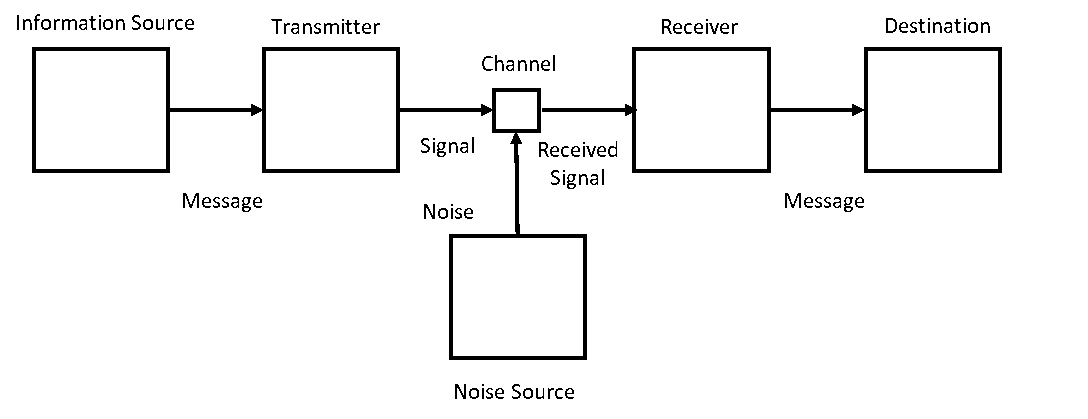}
\caption{Shannon's basic model for communication}
\label{Figure 5}
\end{figure}
%---------------

Systems with an inherent need to adapt and evolve don’t have this luxury of contextual precision, however, and to some extent they need to be context-free and non-Turinian. This is especially true of biological systems, where evidence of genetic redundancy provides a clear sign of nature’s preference for adaptability over communication efficiency. This is seen, for instance, in the structure of DNA, where studies on genetic redundancy are well documented \cite{nowak97}. In such circumstances, informational processes have an inbuilt ability to tolerate error and accommodate semantics simply because they consume context through the information they process, rather than any rigid or imposed embodiment at the communication level. Simply, the purposes for which information has been generated, processed or shared is not lost to any abstract of the processes used to represent and exchange it. This is done by extending strict bipartite models to accommodate variation in both the typing of membership and the number of subsystems allowed to take part. As a result, graph-based models for information representation have become prevalent in nature, with the structure of DNA providing just one example. In such systems, the typing of both nodes and edges is made explicit, either because of some innate property of the physical substrates involved - like chemical composition, genetic structure or higher orders of arrangement - or some rule system embodied in surrounding evolutionary processes. This is not unlike Shannon’s transmitter and receiver model and provides informational context through the position and typing of information \cite{kauffman93} embodied in the overall ontological arrangement of constituent elements at the network level. 

Several sources therefore advocate that a broad-based information theory must be network theoretic, or at least ontological \cite{rovelli14} \cite{w3c13} \cite{graphdb20} \cite{folks20}, whilst still allowing the decoupling of associated semantics from any communications or information persistence frameworks. Vitally, it is this graph-like structure that affords both the representational aspects (nodes) and the transitional nature (edges) of any given problem domain. Others strongly imply support \cite{kauffman93} \cite{west17}. All are further reinforced by at least one view of fundamental physics \cite{rovelli14} which sees reality reduced purely to a realm of possible interactions. Under such an interpretation, any \textit{useful} definition of information would have to fundamentally accept that “all variable aspects of an object (or concept) exist only in relation to other objects (or concepts)” \cite{rovelli14}, so that, “in the world described by quantum mechanics, there is no other way to describe reality than through the relations between physical systems...It isn’t that things enter into relations, but rather relations that ground the notion of ‘thing’” \cite{rovelli14}.

With this as a backdrop, it is worth noting that at least one example of a formally ontological approach based on triples \cite{guha92} has already been implemented at scale using modern information technology infrastructure, in the Semantic Web \cite{w3c13}. Likewise, a much more informal variant of that approach has seen unquestionable success through its adoption by today’s \textit{hashtag} social media movement \cite{folks20} \cite{cohen17}. Both are fundamentally network based, as is the physical architecture of many modern quantum computers \cite{ibmqx20}, which suggests a strong potential for experimental research. Furthermore, there is little controversy over whether quantum entanglement can be used to create triple-like arrangements. For instance, it is well understood that the entanglement model can be extended to create valuable network structures, with quantum teleportation \cite{bennett93} being a recognised and significant stepping stone on that journey - for all that the constraints of Coffman-Kundu-Wootters inequality \cite{coffman2000} \cite{ibm06} must be taken into account. 

\section{Appendix F: Closing Comments}
\label{appendix_f}
\setlength{\parskip}{0pt}

Despite all the challenges, if one accepts that the language of the universe is mathematical and that all of its branches stem from the common root that is set theory, then that has to speak to some inborn propensity for the world to seek commonality and relation. At its most fundamental level then, \textit{useful} reality is not just about chaos, distinguishability and \textit{self}, but some intrinsic ability to share existentially, to contribute some level of \textit{hyper-self }and bring forth structure. This is probably best seen in the fields of chemistry and biology. By definition then, all atoms are chemically inert when isolated, yet still we are the evolutionary outcome of the universe’s avoidance of total segregation. We are literally the children of its interplay and the willingness of the atomic and subatomic worlds to enter into at least some form of meaningful relationship. This is what glues reality together at the most fundamental level known. It is what makes the world informative and useful. It is not so much an essence of information. It is essential to information’s formation in the bringing together of difference and equivalence.

For all that Claude Shannon and his contemporaries more than adequately explained the importance of distinguishability and self in information’s makeup, their lessons are perhaps incomplete. They may well have understood how the \textit{is} of this world tightly binds with physical reality, and in doing so have explained that information itself is a physical construct. But the route they took likely overlooked several subtle insights, in that they did not account for the nature of the primitives in the formation of information or the manner in which they might legitimately associate. Addressing such matters brings implications though, in that severe conflicts arise between what is theoretically plausible and practically \textit{useful}. Indeed, for all that the ideas here outline a novel way to address the challenges of semantics in information theory, they may only serve to point the debate towards the notorious roadblock that is infinity. 

Just as Alan Turing taught that closure provides an antidote to the uselessness of never-ending algorithms, and likewise Max Planck drew a line under the divisibility of space, there appears limits on the usefulness of information. Through the need to constrain language expression, it is concluded that semantics is a finitely quantisable information construct and therefore complementary to existing information theories. Furthermore, as this quantisation occurs in accordance with the laws of entropy, that limits the range of semantic expression available across all useful grammars. This extends all current information theories to include the notion of meaning in a graph- theoretic sense and states that no form of formal quantum semantics can ever describe everything – some semantics are theoretically plausible but can never be meaningfully useful. This is analogous to claiming that an infinite range of unique chemical elements is theoretically possible but accepting that the list of practically available elements is limited by the laws of physics into what we understand today as the periodic table.   

By analysing the error density matrices produced by standard quantum computing measurement approaches, it is considered possible to replace binary encoding with more expressive schemes across both qusyms and their entangled relationships. When applied across entangled qusym pairs this provides a basis to form semantic triples through the joining of quantum corollas. This extends QIT to include graph-theoretic approaches to semantics. Furthermore, by subdividing Hilbert space into finer increments for information encoding, a means is created to both persist and compress formally semantic information in quantum networks, thereby creating a basis from which to develop quantum cache, quantum KR and quantum information management systems.

\vskip 0.2in
\bibliography{towards_a_semantic_info_theory}

\begin{thebibliography}{}

\bibitem[\protect\BCAY{Baroni, Bernardi,\ \BBA\ R.}{Baroni
  et~al.}{2010}]{baroni10}
Baroni, M., Bernardi, R., \BBA\ R., Z. \BBOP2010\BBCP.
\newblock \BBOQ Frege in space: A program for compositional distributional
  semantics\BBCQ\
\newblock {\Bem Linguistic Issues in language technology}, {\Bem 9}, 5--110.

\bibitem[\protect\BCAY{Bennett}{Bennett}{1973}]{bennett73}
Bennett, C. \BBOP1973\BBCP.
\newblock \BBOQ Logical reversibility of computation\BBCQ\
\newblock {\Bem IBM Journal of Research and Development}, {\Bem 17}, 525--532.

\bibitem[\protect\BCAY{Bennett\ \BBA\ DiVincenzo}{Bennett\ \BBA\
  DiVincenzo}{2000}]{bennett2000}
Bennett, C.\BBACOMMA\  \BBA\ DiVincenzo, D. \BBOP2000\BBCP.
\newblock \BBOQ Quantum information and computation\BBCQ\
\newblock {\Bem Nature}, {\Bem 404}, 247--255.

\bibitem[\protect\BCAY{Bennett\ \BBA\ Shor}{Bennett\ \BBA\
  Shor}{1998}]{bennett98}
Bennett, C.\BBACOMMA\  \BBA\ Shor, P. \BBOP1998\BBCP.
\newblock \BBOQ Quantum information theory\BBCQ\
\newblock {\Bem IEEE Transactions on Information Theory}, {\Bem 44},
  2725--2742.

\bibitem[\protect\BCAY{Bennett}{Bennett}{1993}]{bennett93}
Bennett, G. \BBOP1993\BBCP.
\newblock \BBOQ Teleporting an unknown quantum state via dual classical and
  einstein–podolsky–rosen channels\BBCQ\
\newblock {\Bem Physical review Letters}, {\Bem 70}, 1895--1899.

\bibitem[\protect\BCAY{Charniak\ \BBA\ Wilks}{Charniak\ \BBA\
  Wilks}{1976}]{charniak76}
Charniak, E.\BBACOMMA\  \BBA\ Wilks, Y. \BBOP1976\BBCP.
\newblock {\Bem Computational Semantics - An Introduction to Artificial
  Intelligence and Natural Language Comprehension}.
\newblock Amsterdam: North-Holland Publishing Company.

\bibitem[\protect\BCAY{Chierchia}{Chierchia}{1985}]{chierchia85}
Chierchia, G. \BBOP1985\BBCP.
\newblock \BBOQ Formal semantics and the grammar of predication\BBCQ\
\newblock {\Bem Linguistic Inquiry}, {\Bem 16}, 417--443.

\bibitem[\protect\BCAY{Chomsky}{Chomsky}{1959}]{chomsky59}
Chomsky, N. \BBOP1959\BBCP.
\newblock \BBOQ On certain formal properties of grammars\BBCQ\
\newblock {\Bem Information and Control}, {\Bem 2}, 137--167.

\bibitem[\protect\BCAY{Clark\ \BBA\ Pullman}{Clark\ \BBA\
  Pullman}{2007}]{clark07}
Clark, S.\BBACOMMA\  \BBA\ Pullman, S. \BBOP2007\BBCP.
\newblock \BBOQ Combining symbolic and distributional models of meaning\BBCQ\
\newblock In {\Bem AAAI Spring Symposium: Quantum Interaction}, \BPGS\ 52--55.

\bibitem[\protect\BCAY{Coffman, Kundu,\ \BBA\ Wootters}{Coffman
  et~al.}{2000}]{coffman2000}
Coffman, V., Kundu, J., \BBA\ Wootters, E. \BBOP2000\BBCP.
\newblock \BBOQ Distributed entanglement\BBCQ\
\newblock {\Bem Physical Review A}, {\Bem 61}.

\bibitem[\protect\BCAY{Cohen}{Cohen}{2017}]{cohen17}
Cohen, M. \BBOP2017\BBCP.
\newblock \BBOQ Folksonomy: Hashtag\BBCQ\
\newblock
  \url{https://wp.opened.uoguelph.ca/mcohen02/2017/04/04/folksonomy-hashtag}.
\newblock [Online; accessed 09-December-2020].

\bibitem[\protect\BCAY{Cohen\ \BBA\ Widdows}{Cohen\ \BBA\
  Widdows}{2015}]{cohen15}
Cohen, T.\BBACOMMA\  \BBA\ Widdows, D. \BBOP2015\BBCP.
\newblock \BBOQ Embedding probabilities in predication space with hermitian
  holographic reduced representations\BBCQ\
\newblock {\Bem International Symposium on Quantum Interaction}, 245--257.

\bibitem[\protect\BCAY{Cohen, Widdows, Schvaneveldt,\ \BBA\ Rindflesch}{Cohen
  et~al.}{2011}]{cohen11}
Cohen, T., Widdows, D., Schvaneveldt, R., \BBA\ Rindflesch, T. \BBOP2011\BBCP.
\newblock \BBOQ Finding schizophrenia’s prozac:emergent relational similarity
  in predication space\BBCQ\
\newblock In {\Bem Proc 5th International Symposium on Quantum Interactions}.

\bibitem[\protect\BCAY{Cohen, Widdows, Vine, Schvaneveldt,\ \BBA\
  Rindflesch}{Cohen et~al.}{2012}]{cohen12}
Cohen, T., Widdows, D., Vine, L., Schvaneveldt, R., \BBA\ Rindflesch, T.
  \BBOP2012\BBCP.
\newblock \BBOQ Many paths lead to discovery: Analogical retrieval of cancer
  therapies\BBCQ\
\newblock In {\Bem Sixth International Symposium on Quantum Interaction}.

\bibitem[\protect\BCAY{Cyganiak, Wood, Lanthaler, Klyne, Carroll,\ \BBA\
  McBride}{Cyganiak et~al.}{2014}]{cyganiak14}
Cyganiak, R., Wood, D., Lanthaler, M., Klyne, G., Carroll, J., \BBA\ McBride,
  B. \BBOP2014\BBCP.
\newblock \BBOQ {RDF 1.1 Concepts and Abstract Syntax}\BBCQ\
\newblock \BTR, World Wide Web Consortium.
\newblock \url{https://www.w3.org/TR/rdf11-concepts [Online; accessed
  25-August-2020]}.

\bibitem[\protect\BCAY{Darvas}{Darvas}{2005}]{symm05}
Darvas, G. \BBOP2005\BBCP.
\newblock \BBOQ Symmetry, order, entropy and information\BBCQ\
\newblock {\Bem Third Conference on the Foundations of Information Science}.

\bibitem[\protect\BCAY{Davies}{Davies}{2007}]{davies07}
Davies, P. \BBOP2007\BBCP.
\newblock {\Bem The First Law: The Conservation of Energy (in the Four Laws
  that Drive the Universe)}, \BPGS\ 24--45.
\newblock Oxford: Oxford University Press.

\bibitem[\protect\BCAY{Deutsch\ \BBA\ Marletto}{Deutsch\ \BBA\
  Marletto}{2015}]{deutsch15}
Deutsch, D.\BBACOMMA\  \BBA\ Marletto, C. \BBOP2015\BBCP.
\newblock
\newblock \BBOQ Constructor theory of information.proc.r.soc.a471:
  20140540\BBCQ.
\newblock
  \url{https://royalsocietypublishing.org/doi/pdf/10.1098/rspa.2014.0540
  [Online; accessed 06-December-2020]}.

\bibitem[\protect\BCAY{Einstein, Podolsky,\ \BBA\ Rosen}{Einstein
  et~al.}{1935}]{einstein35}
Einstein, A., Podolsky, B., \BBA\ Rosen, N. \BBOP1935\BBCP.
\newblock \BBOQ Can quantum-mechanical description of physical reality be
  considered complete?\BBCQ\
\newblock {\Bem Phys. Rev.}, {\Bem 47}, 777--780.

\bibitem[\protect\BCAY{Fisher}{Fisher}{1994}]{fisher94}
Fisher, Y. \BBOP1994\BBCP.
\newblock \BBOQ Fractal image compression\BBCQ\
\newblock {\Bem Fractals}, {\Bem 2}, 347--361.

\bibitem[\protect\BCAY{Garg, Ikbal, K., Vishwakarma, Karanam,\ \BBA\
  Subramaniam}{Garg et~al.}{2019}]{garg19}
Garg, D., Ikbal, S., K., S., Vishwakarma, H., Karanam, H., \BBA\ Subramaniam,
  L. \BBOP2019\BBCP.
\newblock \BBOQ Quantum embedding of knowledge for reasoning\BBCQ\
\newblock {\Bem Advances in Neural Information Processing Systems}, {\Bem 1},
  1.

\bibitem[\protect\BCAY{Gayler}{Gayler}{2015}]{gayler04}
Gayler, R. \BBOP2015\BBCP.
\newblock
\newblock \BBOQ Vector symbolic architectures answer jackendoff's challenges
  for cognitive neuroscience\BBCQ.
\newblock \url{arXiv:cs/0412059}.

\bibitem[\protect\BCAY{Gerlach\ \BBA\ Stern}{Gerlach\ \BBA\
  Stern}{1922}]{gerlach22}
Gerlach, W.\BBACOMMA\  \BBA\ Stern, O. \BBOP1922\BBCP.
\newblock \BBOQ Der experimentelle nachweis der richtungsquantelung im
  magnetfeld\BBCQ\
\newblock {\Bem Zeitschrift für Physik}, {\Bem 9}, 349--352.

\bibitem[\protect\BCAY{Goodman\ \BBA\ Soni}{Goodman\ \BBA\
  Soni}{2017}]{goodman17}
Goodman, R.\BBACOMMA\  \BBA\ Soni, J. \BBOP2017\BBCP.
\newblock New York: Simon and Schuster.

\bibitem[\protect\BCAY{Goodman\ \BBA\ Soni}{Goodman\ \BBA\
  Soni}{2018}]{goodman18}
Goodman, R.\BBACOMMA\  \BBA\ Soni, J. \BBOP2018\BBCP.
\newblock {\Bem The Bomb in a Mind at Play - How Claude Shannon Invented the
  Information Age}.
\newblock New York: Simon and Schuster.

\bibitem[\protect\BCAY{Guha}{Guha}{1992}]{guha92}
Guha, R. \BBOP1992\BBCP.
\newblock {\Bem Contexts: A Formalization and Some Applications}.
\newblock Stanford: Stanford University.

\bibitem[\protect\BCAY{Géron}{Géron}{2019}]{geron19}
Géron, A. \BBOP2019\BBCP.
\newblock {\Bem End-toEnd Machine Learning Project (in Hands on Machine
  Learning with Scikit-Learn, Keras and TensorFlow; Concepts Tools and
  Techniques to Build Intelligent Systems)}.
\newblock Sevastopol: OReilly Media.

\bibitem[\protect\BCAY{Hartley}{Hartley}{1928}]{hartley28}
Hartley, R. \BBOP1928\BBCP.
\newblock \BBOQ Transmission of information\BBCQ\
\newblock {\Bem Bell System Technical Journal}, {\Bem 7}, 535--536.

\bibitem[\protect\BCAY{Horrocks, Patel-Schneider, Boley, Tabet, Grosof,\ \BBA\
  Dean}{Horrocks et~al.}{2004}]{swrl04}
Horrocks, I., Patel-Schneider, P., Boley, H., Tabet, S., Grosof, B., \BBA\
  Dean, M. \BBOP2004\BBCP.
\newblock \BBOQ Swrl: A semantic web rule language\BBCQ\
\newblock \BTR, World Wide Web Consortium.
\newblock \url{https://www.w3.org/Submission/SWRL [Online; accessed
  06-November-2020]}.

\bibitem[\protect\BCAY{Horrocks\ \BBA\ Ruttenberg}{Horrocks\ \BBA\
  Ruttenberg}{2008}]{owl08}
Horrocks, I.\BBACOMMA\  \BBA\ Ruttenberg, A. \BBOP2008\BBCP.
\newblock \BBOQ {OWL Working Group}\BBCQ\
\newblock \BTR, World Wide Web Consortium.
\newblock \url{https://www.w3.org/2007/OWL/wiki/OWL\_Working\_Group [Online;
  accessed 06-November-2020]}.

\bibitem[\protect\BCAY{{International Business Machines}}{{International
  Business Machines}}{2020}]{ibmqx20}
{International Business Machines} \BBOP2020\BBCP.
\newblock \BBOQ {IBM Quantum Experience}\BBCQ\
\newblock \url{https://quantum-computing.ibm.com}.
\newblock [Online; accessed 27-August-2020].

\bibitem[\protect\BCAY{Jäger\ \BBA\ Rogers}{Jäger\ \BBA\
  Rogers}{2012}]{jager12}
Jäger, G.\BBACOMMA\  \BBA\ Rogers, J. \BBOP2012\BBCP.
\newblock \BBOQ Formal language theory: refining the chomsky hierarchy\BBCQ\
\newblock {\Bem Philos Trans R Soc Lond B Biol Sci}, {\Bem 367}, 1956--1970.

\bibitem[\protect\BCAY{Kanerva}{Kanerva}{2009}]{kanerva09}
Kanerva, P. \BBOP2009\BBCP.
\newblock \BBOQ Hyperdimensional computing: An introduction to computing in
  distributed representation with high-dimensional random vectors\BBCQ\
\newblock {\Bem Cognitive Computation}, {\Bem 1}, 139--159.

\bibitem[\protect\BCAY{Kauffman}{Kauffman}{2000}]{kauffman2000}
Kauffman, S. \BBOP2000\BBCP.
\newblock {\Bem The Nonergodic Universe in Investigations}, \BPGS\ 141--157.
\newblock Oxford: Oxford University Press.

\bibitem[\protect\BCAY{Kauffman}{Kauffman}{2008a}]{kauffman08}
Kauffman, S. \BBOP2008a\BBCP.
\newblock {\Bem The Physicists Rebel}, \BPGS\ 19--30.
\newblock New York: Basic Books.

\bibitem[\protect\BCAY{Kauffman}{Kauffman}{2008b}]{kauffman08_1}
Kauffman, S. \BBOP2008b\BBCP.
\newblock {\Bem {The Nonergodic Universe in Reinventing the Sacred}}, \BPGS\
  120--128.
\newblock New York: Basic Books.

\bibitem[\protect\BCAY{Kauffman}{Kauffman}{1993}]{kauffman93}
Kauffman, S. \BBOP1993\BBCP.
\newblock {\Bem The Origins of Order (in Morphology, Maps and Integrated
  Tissues}, \BPGS\ 537--642.
\newblock {Oxford: Oxford University Press}.

\bibitem[\protect\BCAY{Khrennikov}{Khrennikov}{2010}]{khrennikov10}
Khrennikov, V. \BBOP2010\BBCP.
\newblock {\Bem Ubiquitous Quantum Structure: From Psychology to Finance 2009
  Edition}.
\newblock New York: Springer.

\bibitem[\protect\BCAY{Landauer, Flotz,\ \BBA\ Laham}{Landauer
  et~al.}{1998}]{landauer98}
Landauer, T., Flotz, P., \BBA\ Laham, D. \BBOP1998\BBCP.
\newblock \BBOQ An introduction to latent semantic analysis\BBCQ\
\newblock {\Bem Discourse Processes}, {\Bem 25}, 259--284.

\bibitem[\protect\BCAY{Lehmann}{Lehmann}{1992}]{lehmann92}
Lehmann, F. \BBOP1992\BBCP.
\newblock {\Bem Semantic Networks in Artificial Intelligence}.
\newblock Oxford: Pergamon.

\bibitem[\protect\BCAY{Linden, Popescu,\ \BBA\ Short}{Linden
  et~al.}{2009}]{linden09}
Linden, N., Popescu, S., \BBA\ Short, A. \BBOP2009\BBCP.
\newblock \BBOQ Quantum mechanical evolution towards thermal equilibrium\BBCQ\
\newblock {\Bem Phys. Rev. E}, {\Bem 79}, 61--103.

\bibitem[\protect\BCAY{Luhn}{Luhn}{1958}]{luhn58}
Luhn, H. \BBOP1958\BBCP.
\newblock \BBOQ The automatic creation of literature abstracts\BBCQ\
\newblock {\Bem IBM Journal of Research and Development}.

\bibitem[\protect\BCAY{Luhn}{Luhn}{1959}]{luhn59}
Luhn, H. \BBOP1959\BBCP.
\newblock {\Bem Automatic Derivation of Information Retrieval Encodements from
  Machine Readable Texts}.
\newblock New York: International Business Machines.

\bibitem[\protect\BCAY{Marcolli}{Marcolli}{2015}]{marcolli15}
Marcolli, M. \BBOP2015\BBCP.
\newblock
\newblock \BBOQ Graph grammars\BBCQ.
\newblock \url{http://www.its.caltech.edu/~matilde/GraphGrammarsLing.pdf
  [Online; accessed 23-October-2020]}.

\bibitem[\protect\BCAY{Maron\ \BBA\ Kuhns}{Maron\ \BBA\ Kuhns}{1960}]{maron60}
Maron, M.\BBACOMMA\  \BBA\ Kuhns, J. \BBOP1960\BBCP.
\newblock \BBOQ On relevance, probabilistic indexing and information
  retrieval\BBCQ\
\newblock {\Bem Journal of the ACM}, {\Bem 1}, 1.

\bibitem[\protect\BCAY{Masterman}{Masterman}{2005}]{masterman05}
Masterman, M. \BBOP2005\BBCP.
\newblock {\Bem Language, Cohesion and Form}.
\newblock Cambridge: Cambridge University Press.

\bibitem[\protect\BCAY{McBride\ \BBA\ Brickley}{McBride\ \BBA\
  Brickley}{2004}]{rdfcore04}
McBride, B.\BBACOMMA\  \BBA\ Brickley, D. \BBOP2004\BBCP.
\newblock \BBOQ {RDF Core Working Group (Closed)}\BBCQ\
\newblock \BTR, World Wide Web Consortium.
\newblock \url{https://www.w3.org/2001/sw/RDFCore/ [Online; accessed
  06-November-2020]}.

\bibitem[\protect\BCAY{Mikolov, Chen, Corrado,\ \BBA\ Dean}{Mikolov
  et~al.}{2013}]{mikolov13}
Mikolov, T., Chen, K., Corrado, G., \BBA\ Dean, J. \BBOP2013\BBCP.
\newblock
\newblock \BBOQ Efficient estimation of word representations in vector
  space\BBCQ.
\newblock \url{arXiv:1301.3781}.

\bibitem[\protect\BCAY{Mitchell\ \BBA\ Lapata}{Mitchell\ \BBA\
  Lapata}{2010}]{mitchell10}
Mitchell, J.\BBACOMMA\  \BBA\ Lapata, M. \BBOP2010\BBCP.
\newblock \BBOQ Composition in distributional models of semantics\BBCQ\
\newblock {\Bem Cognitive Science}, {\Bem 34}, 1388--1429.

\bibitem[\protect\BCAY{Müller}{Müller}{2007}]{muller07}
Müller, I. \BBOP2007\BBCP.
\newblock {\Bem A history of Thermodynamics: The Doctrine of Energy and
  Entropy}.
\newblock New York: Springer Science and Business Media.

\bibitem[\protect\BCAY{Nielsen\ \BBA\ Chuang}{Nielsen\ \BBA\
  Chuang}{2000}]{nielsen2000}
Nielsen, M.\BBACOMMA\  \BBA\ Chuang, I. \BBOP2000\BBCP.
\newblock {\Bem Quantum Computation and Quantum Information}.
\newblock Cambridge: Cambridge University Press.

\bibitem[\protect\BCAY{Nowak}{Nowak}{1997}]{nowak97}
Nowak, M. \BBOP1997\BBCP.
\newblock \BBOQ Evolution of genetic redundancy\BBCQ\
\newblock {\Bem Nature}, {\Bem 388}, 167--171.

\bibitem[\protect\BCAY{Nyquist}{Nyquist}{1924}]{nyquist24}
Nyquist, H. \BBOP1924\BBCP.
\newblock \BBOQ Certain factors affecting telegraph speed\BBCQ\
\newblock {\Bem Transactions of the American Institute of Electrical
  Engineers}, {\Bem 43}, 412--422.

\bibitem[\protect\BCAY{Pennington, Socher,\ \BBA\ Manning}{Pennington
  et~al.}{2013}]{pennington14}
Pennington, J., Socher, R., \BBA\ Manning, C. \BBOP2013\BBCP.
\newblock \BBOQ Global vectors for word representation\BBCQ\
\newblock {\Bem EMNLP}.

\bibitem[\protect\BCAY{Penrose}{Penrose}{2004}]{penrose04}
Penrose, R. \BBOP2004\BBCP.
\newblock {\Bem The Big Bang and its thermodynamic legacy}, \BPGS\ 686--734.
\newblock London: Jonathan Cape.

\bibitem[\protect\BCAY{Riegel, Gray, Naweed, Khan, Makondo, Akhalw, Qian,
  Fagin, Barahona, Sharma, Ikbal, Karanam, Ne, A.,\ \BBA\ Srivastava}{Riegel
  et~al.}{2015}]{riegel20}
Riegel, R., Gray, A., Naweed, F., Khan, N., Makondo, N., Akhalw, I., Qian, R.,
  Fagin, F., Barahona, F., Sharma, S., Ikbal, S., Karanam, H., Ne, S., A., L.,
  \BBA\ Srivastava, S. \BBOP2015\BBCP.
\newblock
\newblock \BBOQ Logical neural networks\BBCQ.
\newblock \url{https://arxiv.org/abs/2006.1315 [Online; accessed
  26-February-2021]}.

\bibitem[\protect\BCAY{Riemann}{Riemann}{2008}]{riemann08}
Riemann, P. \BBOP2008\BBCP.
\newblock \BBOQ Foundation of statistical mechanics under experimentally
  realistic conditions\BBCQ\
\newblock {\Bem Phys. Rev. Lett.}, {\Bem 101}, 190--403.

\bibitem[\protect\BCAY{Rolf}{Rolf}{1961}]{rolf61}
Rolf, L. \BBOP1961\BBCP.
\newblock \BBOQ Reversibility and heat generation in the computing
  process\BBCQ\
\newblock {\Bem IBM Journal of Research and Development}, {\Bem 5}, 183--191.

\bibitem[\protect\BCAY{Rovelli}{Rovelli}{2014}]{rovelli14}
Rovelli, C. \BBOP2014\BBCP.
\newblock {\Bem Quanta (in Reality is Not What it Seems)}.
\newblock London: Penguin Books.

\bibitem[\protect\BCAY{Schneeloch, Tison,\ \BBA\ Fanto}{Schneeloch
  et~al.}{2019}]{schneeloch19}
Schneeloch, J., Tison, C., \BBA\ Fanto, M. \BBOP2019\BBCP.
\newblock \BBOQ Quantifying entanglement in a 68-billion-dimensional quantum
  state space\BBCQ\
\newblock {\Bem Nature Communications}, {\Bem 10}, 1--2.

\bibitem[\protect\BCAY{Schrödinger}{Schrödinger}{1935}]{schrodinger35}
Schrödinger, E. \BBOP1935\BBCP.
\newblock \BBOQ Discussion of probability relations between separated
  systems\BBCQ\
\newblock {\Bem Mathematical Proceedings of the Cambridge Philosophical
  Society}, {\Bem 31}, 555--563.

\bibitem[\protect\BCAY{Shannon}{Shannon}{1948}]{shannon48}
Shannon, C. \BBOP1948\BBCP.
\newblock \BBOQ A mathematical theory of communication\BBCQ\
\newblock {\Bem The Bell System Technical Journal}, {\Bem 27}, 379--423.

\bibitem[\protect\BCAY{Shu-Kun}{Shu-Kun}{1996}]{shukun96}
Shu-Kun, L. \BBOP1996\BBCP.
\newblock \BBOQ Correlation of entropy with similarity and symmetry\BBCQ\
\newblock {\Bem Journal of Chemical Information and Computer Sciences}, {\Bem
  36}, 367--376.

\bibitem[\protect\BCAY{Sihare\ \BBA\ Nath}{Sihare\ \BBA\ Nath}{2016}]{sihare16}
Sihare, S.\BBACOMMA\  \BBA\ Nath, V. \BBOP2016\BBCP.
\newblock \BBOQ Application of quantum search algorithms as a web search
  engine\BBCQ\
\newblock {\Bem International Conference on Global Trends in Signal Processing,
  Information Computing and Communication (ICGTSPICC)}.

\bibitem[\protect\BCAY{Simmons}{Simmons}{1963}]{simmons63}
Simmons, R. \BBOP1963\BBCP.
\newblock \BBOQ Synthetic language behavior\BBCQ\
\newblock {\Bem Data Processing Management}, {\Bem 5}, 11--18.

\bibitem[\protect\BCAY{Simmons}{Simmons}{1982}]{simmons82}
Simmons, R. \BBOP1982\BBCP.
\newblock \BBOQ Themes from 1972\BBCQ\
\newblock {\Bem 20th Annual Meeting of the Association for Computational
  Linguistics}.

\bibitem[\protect\BCAY{Smolensky}{Smolensky}{1990}]{smolenksy90}
Smolensky, P. \BBOP1990\BBCP.
\newblock \BBOQ Tensor product variable binding and the representation of
  symbolic structures in connectionist systems\BBCQ\
\newblock {\Bem Artificial Intelligence}, {\Bem 46}, 159--216.

\bibitem[\protect\BCAY{Sparck~Jones\ \BBA\ Kay}{Sparck~Jones\ \BBA\
  Kay}{1973}]{jones73}
Sparck~Jones, K.\BBACOMMA\  \BBA\ Kay, M. \BBOP1973\BBCP.
\newblock {\Bem Linguistics and Information Science}.
\newblock London: Academic Press.

\bibitem[\protect\BCAY{Sparck~Jones\ \BBA\ Willett}{Sparck~Jones\ \BBA\
  Willett}{1997}]{jones97}
Sparck~Jones, K.\BBACOMMA\  \BBA\ Willett, P. \BBOP1997\BBCP.
\newblock {\Bem Readings in Information Retrieval}.
\newblock Burlington: Morgan Kaufmann.

\bibitem[\protect\BCAY{Terhal}{Terhal}{2004}]{ibm06}
Terhal, B.~M. \BBOP2004\BBCP.
\newblock \BBOQ Is entanglement monogamous?\BBCQ\
\newblock {\Bem IBM Journal or Research and Development}, {\Bem 48\/}(1),
  71--78.

\bibitem[\protect\BCAY{Turney}{Turney}{2012}]{turney12}
Turney, P. \BBOP2012\BBCP.
\newblock \BBOQ Domain and function: A dual-space model of semantic relations
  and compositions\BBCQ\
\newblock {\Bem Journal of Artificial Intelligence Research (JAIR)}, {\Bem 44},
  533--585.

\bibitem[\protect\BCAY{van Rijsbergen}{van Rijsbergen}{1995}]{rijsbergen95}
van Rijsbergen, K. \BBOP1995\BBCP.
\newblock
\newblock \BBOQ Automatic text analysis\BBCQ.
\newblock \url{http://www.dcs.gla.ac.uk/Keith/pdf/Chapter2.pdf} [Online;
  accessed 29-October-2020].

\bibitem[\protect\BCAY{van Rijsbergen}{van Rijsbergen}{2004}]{rijsbergen04}
van Rijsbergen, K. \BBOP2004\BBCP.
\newblock {\Bem The Geometry of Information Retrieval}.
\newblock Cambridge: Cambridge University Press.

\bibitem[\protect\BCAY{Wang, Hu, Sanders,\ \BBA\ Ka}{Wang
  et~al.}{2020}]{wang20}
Wang, Y., Hu, Z., Sanders, B., \BBA\ Ka, S. \BBOP2020\BBCP.
\newblock
\newblock \BBOQ Qudits and high-dimensional quantum computing\BBCQ.
\newblock \url{https://arxiv.org/abs/2008.00959}.

\bibitem[\protect\BCAY{West}{West}{2017a}]{west171}
West, G. \BBOP2017a\BBCP.
\newblock {\Bem The Big Picture (in Scale - The Universal Laws of Life and
  Death in Organisms, Cities and Companies)}, \BPGS\ 1--33.
\newblock London: Weidenfield and Nicolson.

\bibitem[\protect\BCAY{West}{West}{2017b}]{west17}
West, G. \BBOP2017b\BBCP.
\newblock {\Bem {Scale - The Universal Laws of Life and Death in Organisms,
  Cities and Companies}}.
\newblock Oxford: Oxford University Press.

\bibitem[\protect\BCAY{Widdows}{Widdows}{2004}]{widdows04}
Widdows, D. \BBOP2004\BBCP.
\newblock {\Bem Geometry and Meaning}.
\newblock Stanford: CLSI.

\bibitem[\protect\BCAY{Widdows}{Widdows}{2008}]{widdows08}
Widdows, D. \BBOP2008\BBCP.
\newblock \BBOQ Semantic vector products: Some initial investigations\BBCQ\
\newblock In {\Bem Proceedings of the Second International Symposium on Quantum
  Interaction}.

\bibitem[\protect\BCAY{{Widdows, D. and Kitto K. and Cohen T.}}{{Widdows, D.
  and Kitto K. and Cohen T.}}{2020}]{widdows20}
{Widdows, D. and Kitto K. and Cohen T.} \BBOP2020\BBCP.
\newblock
\newblock \BBOQ {Quantum Mathematics in Artificial Intelligence}\BBCQ.
\newblock \url{https://arxiv.org/abs/2101.04255} [Online; accessed
  19-January-2020].

\bibitem[\protect\BCAY{{Wikipedia contributors}}{{Wikipedia
  contributors}}{2020a}]{entr20}
{Wikipedia contributors} \BBOP2020a\BBCP.
\newblock \BBOQ Entropy of entanglement --- {W}ikipedia{,} the free
  encyclopedia\BBCQ\
\newblock
  \url{https://en.wikipedia.org/w/index.php?title=Entropy\_of\_entanglement&oldid=1002537445}.
\newblock [Online; accessed 24-August-2020].

\bibitem[\protect\BCAY{{Wikipedia contributors}}{{Wikipedia
  contributors}}{2020b}]{folks20}
{Wikipedia contributors} \BBOP2020b\BBCP.
\newblock \BBOQ Folksonomy --- {Wikipedia}{,} the free encyclopedia\BBCQ\
\newblock \url{https://en.wikipedia.org/wiki/Folksonomy}.
\newblock [Online; accessed 26-August-2020].

\bibitem[\protect\BCAY{{Wikipedia contributors}}{{Wikipedia
  contributors}}{2020c}]{graphdb20}
{Wikipedia contributors} \BBOP2020c\BBCP.
\newblock \BBOQ Graph database --- {Wikipedia}{,} the free encyclopedia\BBCQ\
\newblock
  \url{https://en.wikipedia.org/w/index.php?title=Graph\_database&oldid=1053041111}.
\newblock [Online; accessed 26-August-2020].

\bibitem[\protect\BCAY{Wilks}{Wilks}{2008}]{wilks08}
Wilks, Y. \BBOP2008\BBCP.
\newblock \BBOQ What would a wittgensteinian computational linguistics be
  like?\BBCQ\
\newblock {\Bem AISB 2008 Convention: Communication, Interaction and Social
  Intelligence}.

\bibitem[\protect\BCAY{Wilks}{Wilks}{2019}]{wilks19}
Wilks, Y. \BBOP2019\BBCP.
\newblock {\Bem Transformational Grammars Again (SP series, System Development
  Corporation)}.
\newblock Illinois: University of Illinois.

\bibitem[\protect\BCAY{Wittgenstein}{Wittgenstein}{1933}]{wittgenstein33}
Wittgenstein, L. \BBOP1933\BBCP.
\newblock {\Bem Tractatus Logico-Philosophicus}.
\newblock New York: Harcourt, Brace.

\bibitem[\protect\BCAY{{World Wide Web Consortium}}{{World Wide Web
  Consortium}}{2013}]{w3c13}
{World Wide Web Consortium} \BBOP2013\BBCP.
\newblock \BBOQ {Semantic Web Activity}\BBCQ\
\newblock \BTR, World Wide Web Consortium.
\newblock \url{https://www.w3.org/2001/sw [Online; accessed
  18-September-2020]}.

\bibitem[\protect\BCAY{Zipf}{Zipf}{1959}]{zipf49}
Zipf, H. \BBOP1959\BBCP.
\newblock {\Bem Human Behaviour and the Principle of Least Effort}.
\newblock Cambridge, MA: Addison Wesley.

\end{thebibliography}
\bibliographystyle{theapa}

\end{document}